\documentclass{iopconfser}

\usepackage{bm}
\usepackage{amsmath,amssymb,mathrsfs}
\usepackage{physics}
\usepackage{graphicx}

\begin{document}

\title{Microscopic Investigation of $\alpha$ Clustering for the $\alpha$ Decay Based on the Density-Constrained Frozen Hartree-Fock Calculations}

\author{Shu Yamamura$^{1,*}$ and Kazuyuki Sekizawa$^{1,2,3,\dag}$}

\affil{$^1$Department of Physics, School of Science, Institute of Science Tokyo, Tokyo 152-8550, Japan}
\affil{$^2$Nuclear Physics Division, Center for Computational Sciences, University of Tsukuba, Ibaraki 305-8577, Japan}
\affil{$^3$RIKEN Nishina Center, Saitama 351-0198, Japan}

\email{$^*$yamamura.s.738e@m.isct.ac.jp, $^\dag$sekizawa@phys.sci.isct.ac.jp}

\begin{abstract}
The inter-nucleus potential between an $\alpha$ particle and a daughter nucleus ($^{100}$Sn)
is calculated based on the density-constraint frozen Hartree-Fock (DCFHF) method with a
Skyrme-type energy density functional (EDF). It is shown that the potential has a repulsive
core at short distances with an attractive pocket close to the nuclear surface. We have confirmed that the potential obtained with the Skyrme DCFHF method agrees well with that
obtained with the extended Thomas-Fermi approximation to the Skyrme EDF. Furthermore,
it is demonstrated that the height of the repulsive core decreases by taking into account
melting of the $\alpha$ cluster inside a nucleus. With the aid of the nucleon localization
function, we show that the obtained state in DCFHF clearly manifests localization of spin-up
and spin-down neutrons and protons, suggesting the possible emergence of an $\alpha$-cluster-like structure in the DCFHF wave function.
These results opens up a possibility to investigate $\alpha$ clustering in heavy nuclei within
mean-field calculations.
\end{abstract}

\section{Introduction}

Clustering in atomic nuclei can be regarded as an emergent phenomenon in a many-body system
and has attracted great interest over the years. The most prominent example is the $\alpha$
clustering in relatively light nuclei as well as $\alpha$ decays in heavy systems. Although
the study of the $\alpha$ clustering in light nuclei has long history, the $\alpha$ clustering
in heavy nuclei has not yet been sufficiently clarified. Recently, the generalized relativistic mean-field model  revealed that the $\alpha$ formation occurs at
the surface of Sn isotopes \cite{PhysRevC.89.064321} and the $\alpha$-knockout reaction experiment actually confirmed
this picture \cite{doi:10.1126/science.abe4688}. Concerning the $\alpha$ decays, in the early
20th century, Gamow \cite{Gamow1928}, Gurney and Condon \cite{PhysRev.33.127} successfully
described the decay mechanism. In their theory, an $\alpha$ particle is pre-formed within
the parent nucleus, which periodically collides with the Coulomb barrier, and it is ultimately
emitted by tunneling through the barrier. According to this picture, the decay width contains a product of two factors: one is the preformation probability of the $\alpha$ particle and
the other is the tunneling probability through the Coulomb barrier. In particular, since the
latter factor depends exponentially on the Q value, which determines the kinetic energy of the emitted $\alpha$ particle, the decay width and half-life vary significantly depending on nuclides.

For the calculation of the penetrability, the potential between an $\alpha$ and a daughter
nucleus (which we denote as the $\alpha$-D potential in this article) is assumed. Conventionally,
the double-folding model \cite{Seif_2018}, which is calculated with densities of two nuclei
and phenomenological NN interaction, or the method applying the Skyrme-type energy density
functionals (EDF) \cite{PhysRevC.91.014322} have been used. However, both of them are not
microscopic and sometimes yield potentials with different characters.

The purpose of this study is to evaluate the $\alpha$-D potential microscopically to
elucidate the mechanism of $\alpha$ clustering and decays in various nuclei. Here, we employ
the Hartree-Fock theory with Skyrme-type EDF supplemented with density-constraint technique
and we discuss the $\alpha$-D potential and the $\alpha$ clustering in heavy nuclei.

\section{Theoretical Framework}

\subsection{$\alpha$-D potential}

The $\alpha$-D potential $V(R)$ between an $\alpha$ particle and a daughter nucleus (D)
separated by a relative distance $R$ is defined as
\begin{equation}
    V(R) = E_{\alpha\text{D}}(R) - (E_{\alpha} + E_\text{D}),
\end{equation}
where $E_{\alpha\text{D}}(R)$ denotes the energy of the $\alpha$-D composite system,
and $E_{\alpha} + E_\text{D}$ is the sum of the energies of the isolated $\alpha$ particle
and daughter nucleus.

In this study, $E_{\alpha}$ and $E_\text{D}$ are calculated based on the standard Skyrme HF
theory, while $E_{\alpha\text{D}}(R)$ is calculated using a density-constraint technique,
explained in the next section.

\subsection{Density-constrained frozen Hartree-Fock (DCFHF) method}

The energy $E_{\alpha\text{D}}(R)$ is also calculated with the Skyrme HF approach supplemented
with a density-constraint technique. Namely, we minimize the energy with constraints on the
density distribution to be the same as the sum of the densities of the individual $\alpha$
particle and the daughter nucleus, separated by a distance $R$. To archive this, we employ
the so-called density-constrained frozen Hartree-Fock (DCFHF) method \cite{PhysRevC.95.031601}.
In the DCFHF method, the variation is performed with constraints on the density,
\begin{equation}
    \frac{\delta}{\delta\phi_i^*(\bm{r}\sigma)}\bigl<\Phi\big|
        \hat{H} - \sum_{q=n,p}\int \lambda_q(\bm{r}')\biggl(
        \sum_{j\in q}\delta(\bm{r}-\bm{r}_j)  - n_{\alpha\text{D}}^{(q)}(\bm{r}';R)\biggr) \dd\bm{r}'
    \big|\Phi\bigr> = 0,
    \label{variationDCFHF}
\end{equation}
where $\big|\Phi\bigr>$ is the single Slater determinant, $\lambda_q(\bm{r})$ is the Lagrange multiplier defined at each point in space, and the $n_{\alpha D}^{(q)}(\bm{r};R)$
is the fixed density of the $\alpha$ particle and the daughter nucleus at the distance
$R$,
which is given by
\begin{equation}
    n_{\alpha {\text{D}}}^{(q)}(\bm{r};\bm{R})
    = n_\alpha^{(q)}(\bm{r}-\bm{R}) + n_{\text{D}}^{(q)}(\bm{r}). \label{FD}
\end{equation}
As the result of variation, we obtain density-constraint Skyrme HF equations:
\begin{equation}
\sum_{\sigma'}\Bigl[ \hat{h}_{\sigma\sigma'}^{(q)}(\bm{r})-\lambda_q(\bm{r}) \Bigr]\phi_i(\bm{r}\sigma') = \varepsilon_i\phi_i(\bm{r}\sigma'),
\end{equation}
where $\hat{h}_{\sigma\sigma'}^{(q)}(\bm{r})$ is the single-particle Hamiltonian which
contains various mean-field potentials. Since the mean-field potentials depend
on densities which depend on the solutions $\{\phi_i\}$, the Skyrme HF
equations are solved self-consistently. In practice, the Lagrange multiplier
$\lambda_q(\bm{r})$ is updated iteratively according to the following equation
\cite{PhysRevC.100.024623}:
\begin{align}
    \lambda^{(n)}(\bm{r}) = \lambda^{(n-1)}(\bm{r}) 
    + C\frac{n^{(q)}(\bm{r})-n_{\alpha {\text{D}}}^{(q)}(\bm{r})}{n^{(q)}(\bm{r}) + n_c}
\end{align}
where $C$ and $n_c$ are numerical parameters that control the strength of
the constraint and the stability of convergence. With the
self-consistent converged solution, the energy of the $\alpha$+D-like system,
$E_{\alpha\text{D}}(R)=\langle \Phi(R) | \hat{H} | \Phi(R) \rangle$ is calculated.
 
For comparison, we also calculate $E_{\alpha\text{D}}(R)$ employing the extended
Thomas-Fermi (ETF) approximation \cite{PhysRevC.91.014322}, which partly takes into account the Pauli
blocking effect. In the ETF approximation, the local densities, the kinetic-energy
density $\tau$ and the spin-current density $\bm{J}$, are calculated using
the number density given by Eq.~\eqref{FD}, and the energy is calculated with
those densities as $E_{\alpha\text{D}}^\text{ETF}(R)=E_{\text{Skyrme}}[n(R),\tau^{\text{ETF}}(R),\bm{J}^{\text{ETF}}(R)]$.

\subsection{Nucleon localization function}

To assess the degree of $\alpha$ clustering in the $\alpha$+D-like system obtained
with the DCFHF method, we employ the nucleon localization function (NLF) \cite{PhysRevC.83.034312}. The NLF is defined as
\begin{equation}
\mathcal{C}_{q\sigma}(\bm{r}) = \left[ 1 + \left(\frac{n_{q\sigma}\tau_{q\sigma}-\frac{1}{4}(\nabla n_{q\sigma})^2 - \bm{j}_{q\sigma}^2}{n_{q\sigma}\tau^{\rm{TF}}_{q\sigma}} \right)^2\right]^{-1} \label{NLF}.
\end{equation}
By definition, at any position $\bm{r}$, the NLF takes a value in a range of
[0,1]. When the NLF for $(q\sigma)$ takes a value of 0.5 at position $\bm{r}$
(\textit{i.e.}\ $C_{q\sigma}(\bm{r})=0.5$), it means that particles with the same
$(q\sigma)$ value exist in a similar way as uniform Fermi gas in the vicinity
of position $\bm{r}$. In contrast, if particles with a specific $(q\sigma)$
value are localized—--that is, intuitively, if no other particles with the same $(q\sigma)$ are found around position $\bm{r}$---the NLF reaches unity (\textit{i.e.}, $C_{q\sigma}(\bm{r})=1$).
Therefore, a region where $\mathcal{C}_{p\uparrow}(\bm{r})\simeq
\mathcal{C}_{p\downarrow}(\bm{r})\simeq\mathcal{C}_{n\uparrow}(\bm{r})\simeq
\mathcal{C}_{n\downarrow}(\bm{r})\simeq 1$ could be a signature of $\alpha$-cluster-like localization in the composite system.

\subsection{Numerical details}

We have newly developed a computational code for Skyrme-HF calculations in
three-dimensional Cartesian coordinates with the density-constraint written
in the Julia language. In the calculations presented in this article, a box
of $20\,\text{fm}\times20\,\text{fm}\times30\,\text{fm}$ with a 1-fm mesh is used.
The first and second spatial derivatives are evaluated with the 21-point
finite-difference formulas, where isolated (box) boundary conditions are
adopted at the edges of the box. The Coulomb potential is computed by executing
fast Fourier transform routines. The Skyrme HF equations are solved using
the imaginary-time evolution method. For the parameters in the density-constraint
iterations, we have primarily used $C=5$\,MeV and $n_c=0.05$\,fm$^{-3}$ and make
adjustments around those values when necessary. In this report, we present
preliminary results for the $\mathrm{^{104}Te \to \alpha + ^{100}Sn}$ system
obtained using the SLy4d Skyrme parameter set \cite{Ka-Hae_Kim_1997}.

\section{Results}

\subsection{$\alpha$-D potential}

\begin{figure}[tb]
    \centering
    \includegraphics[width=1.0\linewidth]{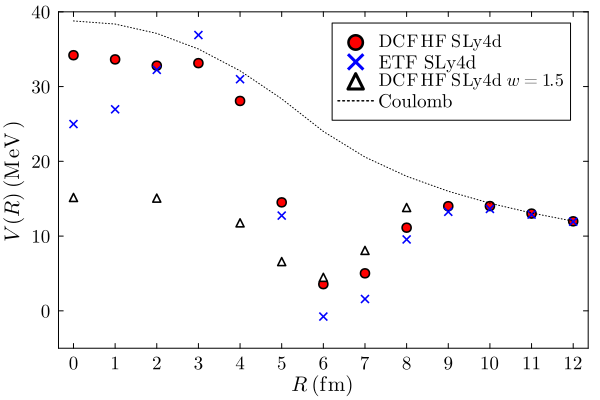}
    \caption{
    The $\alpha$-D potentials, $V(R)$, are plotted as functions of the relative
    distance, $R$. The results obtained with the DCFHF method are shown by
    red filled circles, while those obtained within the ETF approximation are
    shown by blue crosses. The Coulomb potential between a uniformly charged
    sphere and a point $\alpha$ particle is also shown by a dotted line for
    comparison. The black open triangles show the results of DCFHF calculations
    with diffusive $\alpha$-particle density with $w=1.5$\,fm (see texts for details).
    }
    \label{fig:alpha_D_pot_100Sn4He_SLy4d}
\end{figure}

In Fig.~\ref{fig:alpha_D_pot_100Sn4He_SLy4d}, the obtained $\alpha$-D potential $V(R)$
is shown as a function of the relative distance $R$. The results of DCFHF calculations
are shown by red filled circles, while those of ETF calculations are shown by blue crosses.
The Coulomb potential for uniformly charged spheres with radius $R_\text{c}=1.2A^{1/3}$ and a point-charged particle is also shown by a dotted line for comparison.

From the figure, we find that the DCFHF potential exhibits a repulsive core at small $R$
($R\lesssim4$\,fm) with an attractive pocket near $R \simeq 6$\,fm. In a larger $R$
region ($R\gtrsim10$\,fm), it coincides with the Coulomb potential, as it should be.
By comparing the results of the DCFHF and ETF calculations, we find rather small difference,
where the ETF potential becomes a little bit smaller or larger as compared with the
DCFHF result. This fact supports previous studies that calculated the $\alpha$-D
potential using ETF and Skyrme-type EDFs for the evaluation of $\alpha$ decays.
The observation indicates that the repulsive core is mainly caused by the Pauli
effects encoded in the kinetic density. 

To investigate the effects of melting of the $\alpha$ cluster inside the nucleus,
we parametrize the density distribution of the $\alpha$ particle using a Gaussian,
$n_\alpha(\bm{r}-\bm{R})=\mathcal{N}\exp\left(-\frac{(\bm{r}-\bm{R})^2}{2w^2}\right)$,
where $w$ controls the size of the $\alpha$ particle.
In Fig.~\ref{fig:alpha_D_pot_100Sn4He_SLy4d}, the results obtained with the
parametrized $\alpha$ density with $w=1.5$\,fm are shown by open triangles.
We note that the original $\alpha$ particle density roughly corresponds to
a Gaussian of which $w$ is slightly less than 1.3\,fm, thus the $\alpha$ particle is more diffusive
in the $w=1.5$-fm case.

From the figure, we find that the repulsive core at short distances gets substantially
lowered when the melting effect is taken into account, whereas in the region of
intermediate $R$ ($R\simeq6$\,fm), the attractive pocket becomes slightly shallower. This
suggests that while the ``melting" of the $\alpha$ particle is favored in the central
region of the nucleus, the spatial extent comparable to that of the original
$\alpha$ particle in a vacuum is preferred at the nuclear surface.

\subsection{$\alpha$-like localization}

\begin{figure}[tb]
    \centering
    \includegraphics[width=1.0\linewidth]{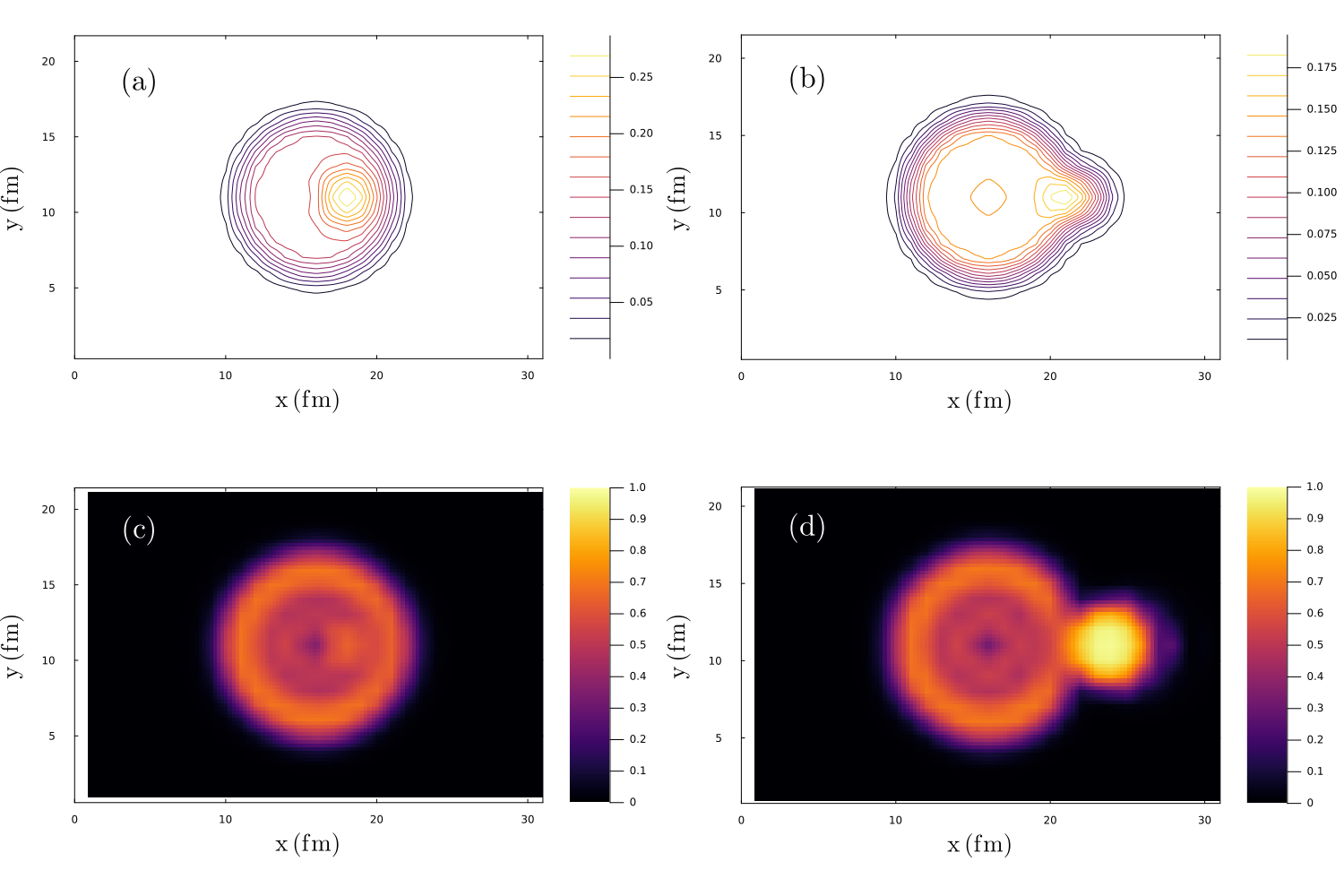}
    \caption{The density [(a) and (b)] and the NLF [(c) and (d)] obtained with
    the DCFHF method. Panels (a) and (c) show the nucleon number density and
    the averaged NLF, respectively, for $R=2~\mathrm{fm}$, where the $\alpha$-D potential
    has the repulsive core. Panels (b) and (d) show the same quantities, but for the
    $R=6~\mathrm{fm}$ case, where the $\alpha$-D potential exhibits the attractive
    pocket.
    }
    \label{fig:density_and_NLF_20260829}
\end{figure}

To investigate the degree of $\alpha$ clustering in DCFHF wave functions, we analyze
the NLF given by the formula \eqref{NLF}. Figures~\ref{fig:density_and_NLF_20260829}(a)
and \ref{fig:density_and_NLF_20260829}(c) show the density and NLF at $R=2\,\mathrm{fm}$, corresponding to the region of the repulsive core of the potential.
Figures~\ref{fig:density_and_NLF_20260829}(b) and \ref{fig:density_and_NLF_20260829}(d)
show the same quantities at $R=6\,\mathrm{fm}$, corresponding to the region of the
attractive pocket of the potential. To visualize the NLF related to the $\alpha$-cluster-like localization in a single panel, we show a geometric mean of $\mathcal{C}_{p\uparrow}$, $\mathcal{C}_{p\downarrow}$, $\mathcal{C}_{n\uparrow}$, and $\mathcal{C}_{n\downarrow}$ in panels
(c) and (d). In the former case shown in panels (a) and (c), the density locally exceeds
the saturation density and no significant localization is observed in the NLF.
On the other hand, in the case of $R=6\,\mathrm{fm}$ shown in panels (b) and (d),
clear localization is observed in the NLF, even though the majority of the region
corresponding to the $\alpha$ particle is embedded in the daughter nucleus.
This observation suggests that, in further studies, it is possible to discuss $\alpha$ clustering
in heavy nuclei based on the Skyrme HF theory that works with a single Slater determinant.

\section{Summary}

As the first step in a microscopic study of the $\alpha$ clustering and decays,
we have calculated the potential between an $\alpha$ particle and a daughter nucleus
microscopically based on the density-constrained frozen Hartree-Fock (DCFHF) method.
The resulting potential exhibits a repulsive core at short distances and an attractive
pocket near the surface, consistent with potentials obtained with the extended Thomas-Fermi approximation. We have also confirmed that the repulsive core diminishes when the
``melting" of the $\alpha$ particle is taken into account by varying its size parameter. Furthermore, the use of the nucleon localization function (NLF) reveals $\alpha$-like localization
of nucleons near the surface in the DCFHF method. These results suggest the applicability of mean-field methods to investigate $\alpha$ clustering and decays
in heavy nuclei.

\section*{Acknowledgments}
This work is supported by JSPS Grant-in-Aid for Scientific Research,
Grant No. JP25H01269.

\bibliographystyle{iopart-num}
\bibliography{main}

\providecommand{\newblock}{}
\begin{thebibliography}{10}
\expandafter\ifx\csname url\endcsname\relax
  \def\url#1{{\tt #1}}\fi
\expandafter\ifx\csname urlprefix\endcsname\relax\def\urlprefix{URL }\fi
\providecommand{\eprint}[2][]{\url{#2}}

\bibitem{PhysRevC.89.064321}
Typel S 2014 {\em Phys. Rev. C\/} {\bf 89}(6) 064321

\bibitem{doi:10.1126/science.abe4688}
Tanaka J, Yang Z, Typel S, Adachi S, Bai S, van Beek P, Beaumel D, Fujikawa Y, Han J, Heil S, Huang S, Inoue A, Jiang Y, Knösel M, Kobayashi N, Kubota Y, Liu W, Lou J, Maeda Y, Matsuda Y, Miki K, Nakamura S, Ogata K, Panin V, Scheit H, Schindler F, Schrock P, Symochko D, Tamii A, Uesaka T, Wagner V, Yoshida K, Zenihiro J and Aumann T 2021 {\em Science\/} {\bf 371} 260--264

\bibitem{Gamow1928}
Gamow G 1928 {\em Zeitschrift für Physik\/} {\bf 51} 204--212

\bibitem{PhysRev.33.127}
Gurney R~W and Condon E~U 1929 {\em Phys. Rev.\/} {\bf 33}(2) 127--140

\bibitem{Seif_2018}
Seif W~M, Abdelhady A~M~H and Adel A 2018 {\em Journal of Physics G: Nuclear and Particle Physics\/} {\bf 45} 115101

\bibitem{PhysRevC.91.014322}
Seif W~M 2015 {\em Phys. Rev. C\/} {\bf 91}(1) 014322

\bibitem{PhysRevC.95.031601}
Simenel C, Umar A~S, Godbey K, Dasgupta M and Hinde D~J 2017 {\em Phys. Rev. C\/} {\bf 95}(3) 031601

\bibitem{PhysRevC.100.024623}
Scamps G and Hashimoto Y 2019 {\em Phys. Rev. C\/} {\bf 100}(2) 024623

\bibitem{PhysRevC.83.034312}
Reinhard P~G, Maruhn J~A, Umar A~S and Oberacker V~E 2011 {\em Phys. Rev. C\/} {\bf 83}(3) 034312

\bibitem{Ka-Hae_Kim_1997}
Kim K~H, Otsuka T and Bonche P 1997 {\em Journal of Physics G: Nuclear and Particle Physics\/} {\bf 23} 1267

\end{thebibliography}

\end{document}